\batchmode
\documentstyle[12pt]{article}
\newcommand{\nc}{\newcommand}
\nc{\renc}{\renewcommand}

\nc{\half}{{\textstyle{1\over2}}}
\nc{\etal}{\mbox{\it et al. }}
\nc{\ie}{{\it i.e.}}
\nc{\eg}{{\it e.g.}}

\renc{\thefootnote}{\arabic{footnote}}
\nc{\capt}[1]{{\bf Figure.} {\small\sl #1}}

\nc{\eqs}[2]{\mbox{Eqs.~(\ref{#1},\,\ref{#2})}}
\nc{\eq}[1]{\mbox{Eq.~(\ref{#1})}}

\nc{\figs}[2]{\mbox{Figs.~(\ref{#1},\,\ref{#2})}}
\nc{\fig}[1]{\mbox{Fig~.(\ref{#1})}}

\nc{\tag}[1]{\label{#1} \marginpar{{\footnotesize #1}}}
\nc{\mtag}[1]{\label{#1} \mbox{\marginpar{{\footnotesize #1}}}}
\renc{\baselinestretch}{1.5}
\newlength{\overeqskip}
\newlength{\undereqskip}
\nc{\be}[1]{\begin{equation} \mbox{$\label{#1}$}}
\nc{\bea}[1]{\begin{eqnarray} \mbox{$\label{#1}$}}
\nc{\Section}[2]{\section{#2}\label{#1}}
\nc{\Bibitem}[1]{\bibitem{#1}}
\nc{\Label}[1]{\label{#1}}

\nc{\eea}{\vspace{\undereqskip}\end{eqnarray}}
\nc{\ee}{\vspace{\undereqskip}\end{equation}}
\nc{\bdm}{\begin{displaymath}}
\nc{\edm}{\end{displaymath}}
\nc{\dpsty}{\displaystyle}
\nc{\bc}{\begin{center}}
\nc{\ec}{\end{center}}
\nc{\ba}{\begin{array}}
\nc{\ea}{\end{array}}
\nc{\bab}{\begin{abstract}}
\nc{\eab}{\end{abstract}}
\nc{\btab}{\begin{tabular}}
\nc{\etab}{\end{tabular}}
\nc{\bit}{\begin{itemize}}
\nc{\eit}{\end{itemize}}
\nc{\ben}{\begin{enumerate}}
\nc{\een}{\end{enumerate}}
\nc{\bfig}{\begin{figure}}
\nc{\efig}{\end{figure}}
\nc{\arreq}{&\!=\!&}
\nc{\arrmi}{&\!-\!&}
\nc{\arrpl}{&\!+\!&}
\nc{\arrap}{&\!\!\!\approx\!\!\!&}
\nc{\non}{\nonumber\\*}
\nc{\align}{\!\!\!\!\!\!\!\!&&}

\def\lsim{\; \raise0.3ex\hbox{$<$\kern-0.75em
      \raise-1.1ex\hbox{$\sim$}}\; }
\def\gsim{\; \raise0.3ex\hbox{$>$\kern-0.75em
      \raise-1.1ex\hbox{$\sim$}}\; }
\nc{\DOT}{\hspace{-0.08in}{\bf .}\hspace{0.1in}}
\nc{\Laada}{\hbox {$\sqcap$ \kern -1em $\sqcup$}}
\nc\loota{{\scriptstyle\sqcap\kern-0.55em\hbox{$\scriptstyle\sqcup$}}}
\nc\Loota{{\sqcap\kern-0.65em\hbox{$\sqcup$}}}
\nc\laada{\Loota}
\nc{\qed}{\hskip 3em \hbox{\BOX} \vskip 2ex}

\nc{\real}{{\rm I \! R}}
\nc{\Z}{{\sf Z \!\!\! Z}}
\nc{\complex}{{\rm C\!\!\! {\sf I}\,\,}}
\def\bigid{\leavevmode\hbox{\small1\kern-3.8pt\normalsize1}}
\def\id{\leavevmode\hbox{\small1\kern-3.3pt\normalsize1}}
\nc{\slask}{\!\!\!/}
\nc{\bis}{{\prime\prime}}
\nc{\pa}{\partial}
\nc{\na}{\nabla}
\nc{\ra}{\rangle}
\nc{\la}{\langle}
\nc{\goto}{\rightarrow}
\nc{\swap}{\leftrightarrow}

\nc{\EE}[1]{ \mbox{$\cdot10^{#1}$} }
\nc{\abs}[1]{\left|#1\right|}
\nc{\at}[2]{\left.#1\right|_{#2}}
\nc{\norm}[1]{\|#1\|}
\nc{\abscut}[2]{\Abs{#1}_{\scriptscriptstyle#2}}
\nc{\vek}[1]{{\rm\bf #1}}
\nc{\integral}[2]{\int\limits_{#1}^{#2}}
\nc{\inv}[1]{\frac{1}{#1}}
\nc{\dd}[2]{{{\partial #1}\over{\partial #2}}}
\nc{\ddd}[2]{{{{\partial}^2 #1}\over{\partial {#2}^2}}}
\nc{\dddd}[3]{{{{\partial}^2 #1}\over
	{\partial #2 \partial #3}}}
\nc{\dder}[2]{{{d #1}\over{d #2}}}
\nc{\ddder}[2]{{{d^2 #1}\over{d {#2}^2}}}
\nc{\dddder}[3]{{d^2 #1}\over
	{d #2 d #3}}
\nc{\dx}[1]{d\,^{#1}x}
\nc{\dy}[1]{d\,^{#1}y}
\nc{\dz}[1]{d\,^{#1}z}
\nc{\dl}[1]{\frac{d\,^{#1}l}{(2\pi)^{#1}}}
\nc{\dk}[1]{\frac{d\,^{#1}k}{(2\pi)^{#1}}}
\nc{\dq}[1]{\frac{d\,^{#1}q}{(2\pi)^{#1}}}

\nc{\cc}{\mbox{$c.c.$ }}
\nc{\hc}{\mbox{$h.c.$ }}
\nc{\cf}{cf.\ }
\nc{\erfc}{{\rm erfc}}
\nc{\Tr}{{\rm Tr\,}}
\nc{\tr}{{\rm tr\,}}
\nc{\pol}{{\rm pol}}
\nc{\sign}{{\rm sign}}
\nc{\bfT}{{\bf T }}

\def\GeV{{\rm\ GeV}}
\def\MeV{{\rm\ MeV}}

\nc{\cA}{{\cal A}}
\nc{\cB}{{\cal B}}
\nc{\cD}{{\cal D}}
\nc{\cE}{{\cal E}}
\nc{\cG}{{\cal G}}
\nc{\cH}{{\cal H}}
\nc{\cL}{{\cal L}}
\nc{\cO}{{\cal O}}
\nc{\cT}{{\cal T}}
\nc{\cN}{{\cal N}}
\nc{\rvac}[1]{|{\cal O}#1\rangle}
\nc{\lvac}[1]{\langle{\cal O}#1|}
\nc{\rvacb}[1]{|{\cal O}_\beta #1\rangle}
\nc{\lvacb}[1]{\langle{\cal O}_\beta #1 |}
\nc{\bb}{\bar{\beta}}
\nc{\bt}{\tilde{\beta}}
\nc{\ctH}{\tilde{\cal H}}
\nc{\chH}{\hat{\cal H}}
\nc{\1}{\aa}
\nc{\2}{\"{a}}
\nc{\3}{\"{o}}
\nc{\4}{\AA}
\nc{\5}{\"{A}}
\nc{\6}{\"{O}}
\nc{\al}{\alpha}
\nc{\g}{\gamma}
\nc{\Del}{\Delta}
\nc{\e}{\epsilon}
\nc{\eps}{\epsilon}
\nc{\lam}{\lambda}
\nc{\om}{\omega}
\nc{\Om}{\Omega}
\nc{\ve}{\varepsilon}
\nc{\mn}{{\mu\nu}}
\nc{\k}{\kappa}
\nc{\vp}{\varphi}

\nc{\advp}[3]{{\it  Adv.\ in\ Phys.\ }{{\bf #1} {(#2)} {#3}}}
\nc{\annp}[3]{{\it  Ann.\ Phys.\ (N.Y.)\ }{{\bf #1} {(#2)} {#3}}}
\nc{\apl}[3]{{\it  Appl. Phys. Lett. }{{\bf #1} {(#2)} {#3}}}
\nc{\apj}[3]{{\it  Ap.\ J.\ }{{\bf #1} {(#2)} {#3}}}
\nc{\apjl}[3]{{\it  Ap.\ J.\ Lett.\ }{{\bf #1} {(#2)} {#3}}}
\nc{\app}[3]{{\it Astropart.\ Phys.\ }{{\bf #1} {(#2)} {#3}}}
\nc{\cmp}[3]{{\it  Comm.\ Math.\ Phys.\ }{{ \bf #1} {(#2)} {#3}}}
\nc{\cqg}[3]{{\it  Class.\ Quant.\ Grav.\ }{{\bf #1} {(#2)} {#3}}}
\nc{\epl}[3]{{\it  Europhys.\ Lett.\ }{{\bf #1} {(#2)} {#3}}}
\nc{\ijmp}[3]{{\it Int.\ J.\ Mod.\ Phys.\ }{{\bf #1} {(#2)} {#3}}}
\nc{\ijtp}[3]{{\it Int.\ J.\ Theor.\ Phys.\ }{{\bf #1} {(#2)} {#3}}}
\nc{\jmp}[3]{{\it  J.\ Math.\ Phys.\ }{{ \bf #1} {(#2)} {#3}}}
\nc{\jpa}[3]{{\it  J.\ Phys.\ A\ }{{\bf #1} {(#2)} {#3}}}
\nc{\jpc}[3]{{\it  J.\ Phys.\ C\ }{{\bf #1} {(#2)} {#3}}}
\nc{\jap}[3]{{\it J.\ Appl.\ Phys.\ }{{\bf #1} {(#2)} {#3}}}
\nc{\jpsj}[3]{{\it J.\ Phys.\ Soc.\ Japan\ }{{\bf #1} {(#2)} {#3}}}
\nc{\lmp}[3]{{\it Lett.\ Math.\ Phys.\ }{{\bf #1} {(#2)} {#3}}}
\nc{\mpl}[3]{{\it  Mod.\ Phys.\ Lett.\ }{{\bf #1} {(#2)} {#3}}}
\nc{\ncim}[3]{{\it  Nuov.\ Cim.\ }{{\bf #1} {(#2)} {#3}}}
\nc{\np}[3]{{\it  Nucl.\ Phys.\ }{{\bf #1} {(#2)} {#3}}}
\nc{\npps}[3]{{\it  Nucl.\ Phys.\ Proc.\ Suppl.\ }{{\bf #1} {(#2)} {#3}}}
\nc{\pr}[3]{{\it Phys.\ Rev.\ }{{\bf #1} {(#2)} {#3}}}
\nc{\pra}[3]{{\it  Phys.\ Rev.\ A\ }{{\bf #1} {(#2)} {#3}}}
\nc{\prb}[3]{{\it  Phys.\ Rev.\ B\ }{{{\bf #1} {(#2)} {#3}}}}
\nc{\prc}[3]{{\it  Phys.\ Rev.\ C\ }{{\bf #1} {(#2)} {#3}}}
\nc{\prd}[3]{{\it  Phys.\ Rev.\ D\ }{{\bf #1} {(#2)} {#3}}}
\nc{\prl}[3]{{\it Phys.\ Rev.\ Lett.\ }{{\bf #1} {(#2)} {#3}}}
\nc{\pl}[3]{{\it  Phys.\ Lett.\ }{{\bf #1} {(#2)} {#3}}}
\nc{\prep}[3]{{\it Phys.\ Rep.\ }{{\bf #1} {(#2)} {#3}}}
\nc{\prsl}[3]{{\it Proc.\ R.\ Soc.\ London\ }{{\bf #1} {(#2)} {#3}}}
\nc{\ptp}[3]{{\it  Prog.\ Theor.\ Phys.\ }{{\bf #1} {(#2)} {#3}}}
\nc{\ptps}[3]{{\it  Prog\ Theor.\ Phys.\ suppl.\ }{{\bf #1} {(#2)} {#3}}}
\nc{\physa}[3]{{\it  Physica\ A\ }{{\bf #1} {(#2)} {#3}}}
\nc{\physb}[3]{{\it  Physica\ B\ }{{\bf #1} {(#2)} {#3}}}
\nc{\phys}[3]{{\it Physica\ }{{\bf #1} {(#2)} {#3}}}
\nc{\rmp}[3]{{\it  Rev.\ Mod.\ Phys.\ }{{\bf #1} {(#2)} {#3}}}
\nc{\rpp}[3]{{\it Rep.\ Prog.\ Phys.\ }{{\bf #1} {(#2)} {#3}}}
\nc{\sjnp}[3]{{\it Sov.\ J.\ Nucl.\ Phys.\ }{{\bf #1} {(#2)} {#3}}}
\nc{\spjetp}[3]{{\it Sov.\ Phys.\ JETP\ }{{\bf #1} {(#2)} {#3}}}
\nc{\yf}[3]{{\it Yad.\ Fiz.\ }{{\bf #1} {(#2)} {#3}}}
\nc{\zetp}[3]{{\it Zh.\ Eksp.\ Teor.\ Fiz.\  }{{\bf #1}  {(#2)} {#3}}}
\nc{\zp}[3]{{\it Z.\ Phys.\ }{{\bf #1} {(#2)} {#3}}}
\nc{\ibid}[3]{{\sl ibid.\ }{{\bf #1} {#2} {#3}}}
\nc{\rf}[1]{(\ref{#1})}
\nc{\nn}{\nonumber \\*}
\nc{\bfB}{\bf{B}}
\nc{\bfv}{\bf{v}}
\nc{\bfx}{\bf{x}}
\nc{\bfy}{\bf{y}}
\nc{\vx}{\vec{x}}
\nc{\vy}{\vec{y}}
\nc{\oB}{\overline{B}}
\nc{\oI}{\overline{I}}
\nc{\oR}{\overline{R}}
\nc{\rar}{\rightarrow}
\nc{\ti}{\times}
\nc{\slsh}{\hskip-5pt/}
\nc{\sm}{Standard~Model~}
\nc{\MP}{M_{\rm Pl}}
\nc{\tp}{t_{\rm Pl}}
\nc{\ave}{\bar{E}}

\nc{\eff}{{\rm eff}}
\nc{\kk}{\vek{k}}
\nc{\pp}{{\rm p}}
\nc{\ga}{g_{a\gamma}}
\nc{\vv}{\\}
\nc{\eee}{{\bf E}}
\nc{\bbb}{{\bf B}}
\nc{\qcd}{T_{\rm QCD}}
\nc{\G}{\rm \ G}
\def\vec#1{{\bf #1}}

\def\lae{\;^{<}_{\sim} \;} \def\gae{\; ^{>}_{\sim} \;} 

\def\ell{e^{c}LL}

\begin{document}
{\title{\vskip-2truecm{\hfill {{\small \\
	\hfill \\
	}}\vskip 1truecm}
{\LARGE Reheating Temperature and Inflaton Mass Bounds from Thermalization After Inflation}}
{\author{
{\sc  John McDonald$^{1}$}\\
{\sl\small Department of Physics and Astronomy,
University of Glasgow, Glasgow G12 8QQ, SCOTLAND}
}
\maketitle
\begin{abstract}
\noindent

              We consider the conditions for the decay products of perturbative inflaton
 decay to thermalize. The importance of  considering the full spectrum of inflaton
 decay products in the thermalization process is emphasized.
It is shown that the delay between the end of inflaton 
decay and thermalization allows the thermal gravitino upper bound on the reheating
 temperature to be raised from $10^{8} \GeV$ to as much as $10^{12} \GeV$ in
 realistic inflation models. 
Requiring that thermalization occurs before nucleosynthesis imposes an upper bound 
on 
the inflaton mass as a function of the reheating temperature, $m_{S} \lae 10^{10} 
(T_{R}/1 \GeV)^{7/9} \GeV$. It is also shown that even in realistic inflation models
 with relatively large reheating temperatures, it is non-trivial to have thermalization
 before the electroweak phase transition temperature.  Therefore the thermal history of
 the Universe is very sensitive to details of the inflation model.              
\end{abstract}
\vfil
\footnoterule
{\small $^1$mcdonald@physics.gla.ac.uk}

\thispagestyle{empty}
\newpage
\setcounter{page}{1}

\section{Introduction}

              Reheating is a fundamental process in early Universe cosmology \cite{eu}, in which the energy density in a coherently oscillating inflaton field 
is converted to thermalized relativistic particles. Originally it was believed that this occurred simply by the perturbative 
decay of the individual scalar particles in the corresponding Bose condensate of inflatons \cite{eu,ptherm}, 
but in recent times it has become clear that the process can be considerably more complicated, with 
non-perturbative processes such as parametric resonance \cite{preh} and quantum 
creation of fermions \cite{dgtr0,linde,dgtr} playing a significant role. Nevertheless, 
in many models, even if there is an early stage of preheating, 
the latter stage of reheating is dominated by the perturbative decay 
of the remaining inflaton energy density. An important issue is then 
the thermalization of the inflaton decay products. Since the 
inflaton can be a very massive particle, as heavy as 
$10^{15} \GeV$ in some typical inflation models \cite{dti,infm}, it is 
not obvious that its highly energetic decay products will 
thermalize rapidly. In this paper we will consider 
the conditions under which complete thermalization of the inflaton decay products 
 occurs\begin{footnote}{An earlier discussion of the thermalization of inflaton decay products is given in \cite{olive}.}\end{footnote}. 
We will see that the results are particularly important in SUSY inflation models, 
allowing the thermal gravitino upper bound on the reheating temperature 
\cite{thermg,bbn} to be substantially increased. In addition, we will show that
 thermalization can occur at low temperatures even in realistic inflation models, as low
 as the electroweak phase transition temperature or less. We will also calculate the
 upper bounds on the inflaton mass following from the requirement of thermalization
 before nucleosynthesis \cite{eu,bbn}.
 
\section{Thermalization of Inflaton Decay Products}

       There are two distinct processes involved in thermalization of the decay products, 
which we shall refer to as "self-thermalization" and "catalysed thermalization". 

         In order to avoid confusion, we first define what we mean by the "thermalization
temperature" and the "reheating temperature". We will define radiation as 
a background of relativistic particles. The thermalization temperature, $T_{th}$, 
refers to the temperature of the radiation when the relativistic particles can scatter
rapidly enough relative to the expansion rate of the Universe to come into thermal equilibrium. The reheating temperature, $T_{R}$, is defined as the temperature the 
radiation {\it would} have at the time when the Universe becomes radiation dominated 
{\it if} it were in thermal equilibrium. This is the conventional reheating temperature 
of inflation models ($T_{R} \approx (M_{Pl} \Gamma_{d})^{1/2}$, where $\Gamma_{d}$ is the inflaton decay rate), which usually assume that thermalization of the
 inflaton decay products is instantaneous. 
\newline{\bf (i) Self-Thermalization}

        The energy density of the decaying inflaton field, $S$, is given by 
\be{e0}  \rho_{S} = \left(\frac{a_{o}}{a}\right)^{3}\rho_{So} e^{-\Gamma_{d}t}   ~,\ee 
where $a$ is the scale factor. (We are assuming here that a single decaying inflaton field is
 the source of the thermal energy.)
Thus most of the energy density in the inflaton field decays when $H \approx \Gamma_{d}$ (where $H = 2/3t$ is the expansion rate during inflaton matter domination), just before the
 Universe becomes dominated by relativistic particles. 
Therefore the apparent condition for the inflaton 
decay products to thermalize is that these decay products should
thermalize by scattering from each other. (This is the condition considered in \cite{olive}.) We refer to this process, 
the thermalization of decay products produced during an interval $\delta t \approx H^{-1}$ by 
scattering from each other, as self-thermalization. 
However, we will see later that this condition for the thermalization of the radiation 
background is incorrect; there are also much lower energy 
particles in the spectrum of decay products, coming from
the red-shifted decay products of earlier inflaton decays, 
which play a crucial role in the thermalization process. 

          We first derive an upper bound on the inflaton mass from self-thermalization of the
 decay products produced during $\delta t \approx H^{-1}$ at the end of inflaton decay. The
 initial energy of the decay products will be of the order of the inflaton mass, $m_{S}$. 
The condition for the thermalization of these decay products by scattering from each other  is
 then 
\be{r1}    \Delta n(H) \sigma_{sc}(H) \gae N_{sc}H      ~,\ee
where $N_{sc}$ is the number of scatterings required to fully thermalize the energy;
 typically $N_{sc}\lae 10$. $\Delta n$ is the number of decay products at $H$ which 
were produced in a time $\delta t \approx H_{R}^{-1}$ at $H_{R}$, where 
$H_{R}$ is the expansion rate when the Universe becomes radiation dominated. 
This is given by 
\be{r2}    \Delta n (H)  \approx \left(\frac{a_{H_{R}}}{a_{H}}\right)^{3} 
\Gamma_{d} H_{R}^{-1} n_{S}     ~,\ee
where $n_{S} \approx \rho_{S}(H_{R})/m_{S}$ is the 
number of inflatons remaining in the condensate at $H_{R}$, $a_{H}$ is the scale
 factor at $H$ and 
$\Gamma_{d}$ is the inflaton decay rate, given by 
\be{r3}  \Gamma_{d} = \frac{k_{T_{R}} T_{R}^{2}}{M_{Pl}} \;;\;\;\;\;\;\; k_{T} = \left(\frac{4 \pi^{3} g(T)}{45}\right)^{1/2}   ~,\ee
where $T_{R}$ is the conventional 
reheating temperature and $g(T)$ is the number of relativistic degrees of freedom. The
 scattering rate at $H$ for relativistic particles of initial energy $E 
\approx m_{S}$ at $H_{R}$ is given by
\be{r4} \sigma_{sc} \approx \frac{\alpha^{2}}{E^{2}}\approx  \left(\frac{a}{a_{R}}\right)^{2} 
\frac{\alpha^{2}}{m_{S}^{2}}     ~,\ee
where $\alpha = g^{2}/4\pi$ corresponds to the gauge or Yukawa couplings. (For
 now we will consider massless decay products.) This assumes that 2 $\rightarrow$ 2
 particle scattering processes can produce final state particles which subsequently
 rapidly decay, so increasing the number density and decreasing the average energy of
 the particles in the ensemble; otherwise we should consider processes such as 2
 $\rightarrow$ 4 particles, with a correspondingly smaller $\alpha$. 
Thus the condition for complete self-thermalization at a temperature $T \; < T_{R}$ is 
\be{r5} m_{S} \lae \left(\frac{T_{R}}{T}\right)^{1/3} 
 \left(\frac{3 M_{Pl} k_{T} \alpha^{2}}{8 \pi N_{sc}}\right)^{1/3} T_{R}^{2/3} \equiv m_{self}
~.\ee
Numerically we find
\be{r5a}  m_{self} = 2.9 \times 10^{7} \; \alpha^{2/3} N_{sc}^{-1/3} \left(\frac{1 \MeV}{T}\right)^{1/3} 
\left(\frac{T_{R}}{1 \GeV}\right)  \GeV     ~,\ee
where we have used $k_{T} \approx 17$ ($g(T) \approx 100$). 
\newline{\bf (ii) Catalysed Thermalization} 

           The naive approach to thermalization considers only the self-thermalization of the
 decay products produced at the end of reheating. In fact, 
there will also be red-shifted decay products from earlier inflaton decays. If these have 
red-shifted sufficiently, their scattering rate ($\propto E^{-2}$) can become large and so they
 can self-thermalize, transferring their energy density from 
a small number of high energy particles to a larger number of low energy particles.
 These can then act as targets for 
higher energy particles in the energy spectrum to scatter from and thermalize,
 with the process continuing until all the decay products are thermalized. 
We refer to this process as 
"catalysed thermalization". The conditions for catalysed thermalization to occur are then that  
(i) there are particles in the energy spectrum of decay 
products of sufficiently low energy as to be able to self-thermalize and so provide a "seed"
 for
catalysed thermalization and (ii) that catalysed thermalization can then thermalize the whole
 spectrum of decay products in a time $\lae H^{-1}$.

     In order to discuss catalysed thermalization, we need the spectrum of decay products, 
$dn(E,T)/dE$, at $T$. Inflaton decay during a time $\delta t_{i} \approx H^{-1}_{i}$ at
 $H_{i}$ contributes
a number density at $H$ given by 
\be{s6} dn(H, H_{i}) \approx \left(\frac{a_{H_{i}}}{a_{H}}\right)^{3} 
\Gamma_{d} H_{i}^{-1} \frac{\rho_{S}(H_{i})}{m_{S}}    ~.\ee
The energy of the decay products red-shifts to 
\be{s7} E = \left(\frac{a_{H_{i}}}{a_{H}}\right) m_{S}  \equiv 
\left(\frac{H_{R}}{H_{i}}\right)^{2/3} \left(\frac{H}{H_{R}}\right)^{1/2} m_{S} ~,\ee
for $H < H_{R}$. In the following, it will be sufficient to consider the spectrum at $H < H_{R}$, since the weakest bounds generally correspond to both the largest red-shift
 of the decay products and the smallest value of $H$. Thus we find
\be{s8} dn(H,H_{i}) = \left(\frac{T}{T_{R}}\right)^{3/2} \frac{\rho_{S}(H_{R})}{m_{S}} 
\left(\frac{E}{m_{S}}\right)^{3/2}     ~.\ee
The change in energy at $H$ of the decay products produced at $H_{i}$ in a time $\delta
 t_{i} \approx H_{i}^{-1}$ is $\delta E \approx 2 E/3$. Thus
\be{s9}    \frac{dn}{dE} \approx \frac{3}{2} \left(\frac{T}{T_{R}}\right)^{3/2} 
\frac{\rho_{S}(H_{R})}{m_{S}} \frac{E^{1/2}}{m_{S}^{3/2}}      ~,\ee
for $ H< H_{R}$.
An important point in what follows is that this spectrum has a low energy cut-off,
 at $E_{min}$, corresponding the the inflaton decay products produced at the earliest time,
 immediately after the end of inflation at $H = H_{I}$, 
\be{s10} E_{min} = \left(\frac{H_{R}}{H_{I}}\right)^{2/3} \left(\frac{H}{H_{R}}\right)^{1/2} 
m_{S} \equiv 
\left(\frac{T}{T_{R}}\right) \left(\frac{k_{T_{R}}T_{R}^{2}}{M_{Pl} H_{I}}\right)^{2/3} m_{S}  
   ~.\ee  
The condition for a self-thermalized seed to exist at $H < H_{R}$ is then that, for some energy  $E_{c} > E_{min}$, self-thermalization of the decay products can occur
for all $E$ up to $E_{c}$,
\be{s11}  \frac{dn}{dE} \frac{\alpha^{2}}{E}  \gae N_{sc}H  \;,\;\;\;\;\; \forall E \lae E_{c} ~.\ee 
Using \eq{s9}, we find that $E_{c}$ is given by 
\be{s12} E_{c} = \frac{9 \pi g(T_{R}) \alpha^{4} M_{Pl}^{2} T_{R}^{5}}{320 m_{S}^{5} N_{sc}^{2} T}     ~.\ee
A self-thermalized seed will therefore exist if $E_{min}$ is less than $E_{c}$ at the
 smallest value of $T$, 
which imposes an upper bound on the inflaton mass
\be{s13}  m_{S} \lae \left(\frac{T_{R}}{T}\right)^{1/3} 
\left( \frac{9 \pi g(T_{R}) M_{Pl}^{8/3} \alpha^{4}}{320 k_{T_{R}}^{2/3} N_{sc}^{2}} \right)^{1/6} 
T_{R}^{4/9} H_{I}^{1/9}  \equiv m_{seed}   ~, \ee
where $ T < T_{R}$. Numerically we find
\be{s14} m_{seed} = 8.8 \times 10^{10} \alpha^{2/3} N_{sc}^{-1/3}
\left(\frac{T_{R}}{1 \GeV}\right)^{7/9} 
\left(\frac{1 \MeV}{T}\right)^{1/3} 
\left(\frac{H_{I}}{10^{13}\GeV}\right)^{1/9}  \GeV    ~.\ee
If this is satisfied, then catalysed thermalization can 
thermalize the spectrum for all $E$ such that
\be{s15} \left(\frac{dn}{dE} E^{2}\right)^{3/4} \frac{\alpha^{2}}{E_{cm}^{2}} \gae N_{sc}H  ~.\ee
$E_{cm} =  \sqrt{EE_{th}}$ is the centre of mass energy for scattering
 between an inflaton decay product of energy $E$ and a thermalized particle of energy
 $E_{th}\approx \rho_{E}^{1/4}$, where $\rho_{E} \approx Edn/dE$ is the
 energy density in particles of energy $E$ to $2E$, and we have taken the number density of
 thermalized particles to be $\approx \rho_{E}^{3/4}$. 
The condition for catalysed thermalization is then that 
\be{s15a} \frac{dn}{dE} \gae \left(\frac{N_{sc}H}{\alpha^{2}}\right)^{2} ~.\ee
Since $dn/dE$ increases with $E$, 
the condition for catalysed thermalization will be satisfied for all $E \gae E_{c}$ so long as
 it is satisfied at $E_{c}$, which requires that $E_{th}(E) < E_{c}$ at 
$E = E_{c}$. This is true so long as  
\be{s16} m_{S} \lae 0.6 \alpha^{6/5} N_{sc}^{-3/5} g(T_{R})^{1/10} \left(\frac{T_{R} M_{Pl}^{3/5}}{T^{3/5}}\right)
\equiv m_{cat}     ~.\ee  
Numerically we find 
\be{s17} m_{cat} \approx 1.7 \times 10^{13} \alpha^{6/5} N_{sc}^{-3/5} 
\left(\frac{T_{R}}{1 \GeV}\right)
\left(\frac{1 \MeV}{T}\right)^{3/5}    \GeV   ~.\ee 
As this is a much weaker upper bound than $m_{seed}$, the 
upper bound from catalysed thermalization is generally given by $m_{S} \lae
 m_{seed}$. The upper bound from catalysed thermalization is considerably weaker
 than the naive upper bound based on self-thermalization, \eq{r5a}, typically by two to
 three orders of magnitude. 

\section{Consequences for Thermal Gravitinos, Nucleosynthesis and the Electroweak Transition}

        So far we have not considered a specific inflation model, so our results are valid 
for both SUSY and non-SUSY models. In general, there is a lower bound on the
 thermalization temperature from nucleosynthesis, $T_{th} \gae 1 \MeV$  \cite{bbn}. In
 addition, 
in SUSY inflation models there is an upper bound from requiring that gravitinos are 
not produced thermally, $T_{th} \lae 10^{8-9} \GeV$ for gravitino masses in the range
 $100-500 \GeV$ \cite{thermg,bbn}.   Usually  it is assumed that the inflaton decay
 products thermalize instantaneously, 
so that $T_{th}$ is identified with $T_{R}$. However, this depends on the inflaton mass. 
Perhaps the most interesting consequence of this is that the thermal gravitino 
upper bound on $T_{R}$ can be considerably relaxed in realistic SUSY inflation models. 
To see this, we note that thermal gravitinos can only be generated at $T \lae T_{th}$. Thus
 the thermal gravitino upper bound should be $T_{th} \lae 10^{8-9} \GeV$. If 
$m_{S} \gae m_{seed}$ when $T \approx 10^{8-9}\GeV$, then 
thermalization will occur safely below the thermal gravitino upper bound for the
 corresponding value of $T_{R}$. In Table 1 we give values of $m_{seed}$ as 
a function of $T_{R}$ for the case $T = 10^{8} \GeV$. (The 
values of $m_{seed}$ for $T = 10^{9} \GeV$ are given by multiplying the values in Table
 1 by 0.46.)
From this we see that for inflaton masses in the range $10^{15-16} \GeV$, as would be
 expected, for example, in D-term inflation models\begin{footnote}{The inflaton mass in D-term inflation models is given by 
$m_{S} = \lambda \xi$, where $\lambda$ is the Yukawa coupling of the inflation
sector fields and the microwave backgoriund implies that $\xi \approx 7 \times 10^{15} \GeV$  \cite{dti,em}}\end{footnote}, the upper bound 
on the reheating temperature $T_{R}$ is $10^{10-12} \GeV$. Given the importance of the
 thermal gravitino upper bound as a constraint on inflation models, this weakening of the
 upper bound on $T_{R}$ is significant. 

\begin{center} {\bf Table 1. Inflaton Mass Lower Bounds 
from Thermal Gravitino Non-production.} \end{center} 
\begin{center}
\begin{tabular}{|c|c|}          \hline
$T_{R}$ & $m_{seed}/(\alpha^{2/3} N_{sc}^{-1/3} 
\left(\frac{H_ {I}}{10^{13} \GeV}\right)^{1/9})$  \\ \hline
$10^{8} \GeV$ & $3.2 \times 10^{13} \GeV$ 
 \\
$10^{9} \GeV$ & $1.9 \times 10^{14}\GeV$ 
 \\
$10^{10} \GeV$ & $1.1 \times 10^{15}\GeV$ 
 \\
$10^{11} \GeV$ & $6.9 \times 10^{15}\GeV$ 
 \\
$10^{12} \GeV$ & $4.1 \times 10^{16}\GeV$ 
 \\
$10^{13} \GeV$ & $2.5 \times 10^{17}\GeV$ 
 \\
$10^{14} \GeV$ & $1.5 \times 10^{18}\GeV$ 
 \\
$10^{15} \GeV$ & $8.9 \times 10^{18}\GeV$ 
 \\
\hline
\end{tabular}
\end{center} 

          The nucleosynthesis lower bound on $T_{th}$ imposes upper bounds on the inflaton
 mass. These bounds are relatively weak for large $T_{R}$, but for smaller values of 
$T_{R}$ they can be significant. In the context of SUSY models there have been some
 recent motivations for 
considering low reheating temperatures. One is from Affleck-Dine baryogenesis
 \cite{ad,drt,jrev}. 
For the lowest dimension R-parity conserving flat directions of the MSSM scalar potential,
 those with dimension $ d = 4$ and 6 (where the dimension refers to the non-renormalizable
 superpotential terms responsible for lifting the flat directions \cite{drt,jrev,nd1}), the
 observed baryon asymmetry requires that the reheating temperature is approximately
 $10^{7} \GeV$ 
and $1 \GeV$ respectively \cite{jrev,nd1}. So $T_{R} \approx 1 \GeV$ is one favoured
 possibility 
if the baryon asymmetry originates via the Affleck-Dine mechanism. Another motivation 
for low reheating temperatures is the possibility that large non-thermal gravitino densities 
are created by the oscillating inflaton field at the end of inflation \cite{dgtr}. Although the
 resulting upper bound on the reheating temperature is sensitive to the details of the 
inflation model, 
there are indications that, for mass scales typical of inflation models, the upper bound is
 likely to be\begin{footnote}{This estimate is based on a model with a single chiral
 superfield, so that the spin-1/2 components of the gravitino are effectively the inflatino
 \cite{dgtr}. It remains to be seen 
whether this remains true in more realistic models.}\end{footnote} $T_{R} \lae 10^{3} \GeV$ \cite{dgtr}. 

\begin{center} {\bf  Table 2. Inflaton Mass Upper Bounds vs. Thermalization Temperature} \end{center} 
\begin{center}
\begin{tabular}{|c|c|c|c|}          \hline
$T_{R}$ & $T_{th}$ & $m_{self}/\alpha^{2/3} $  & $m_{seed} / (\alpha^{2/3} N_{sc}^{-1/3} \left(\frac{H_ {I}}{10^{13} \GeV}\right)^{1/9})$ \\ \hline
$1 \GeV$  & $1 \MeV$ & $2.9 \times 10^{7} \GeV $ &  $9.2 \times 10^{10} \GeV $ 
 \\
$1 \GeV$  & $1 \GeV$ & $2.9 \times 10^{6} \GeV $ &  $9.2 \times 10^{9} \GeV $ 
 \\ \hline
$10^{3} \GeV$  & $1 \MeV$ & $2.9 \times 10^{10} \GeV $ &  $2.0 \times 10^{13} \GeV $ 
 \\
$10^{3} \GeV$  & $10^{2} \GeV$ & $6.3 \times 10^{8} \GeV $ &  $4.3 \times 10^{11} \GeV $ 
 \\
$10^{3} \GeV$  & $10^{3} \GeV$ & $2.9 \times 10^{8} \GeV $ &  $2.0 \times 10^{11} \GeV $ 
 \\ \hline
$10^{8} \GeV$  & $1 \MeV$ & $2.9 \times 10^{15} \GeV $ &  $1.5 \times 10^{17} \GeV $ 
 \\
$10^{8} \GeV$  & $10^{2} \GeV$ & $6.3 \times 10^{13} \GeV $ &  $3.2 \times 10^{15} \GeV $ 
 \\
$10^{8} \GeV$  & $10^{8} \GeV$ & $6.3 \times 10^{11} \GeV $ &  $3.2 \times 10^{13} \GeV $ 
 \\
\hline
\end{tabular}
\end{center} 

            In Table 2 we give the upper bound on the inflaton mass as a function of the
 thermalization temperature for 
$T_{R} = 1 \GeV,\; 10^{3} \GeV,$  and $10^{8} \GeV$. 
For $T_{R} \approx 1 \GeV$ the nucleosynthesis upper bound 
from catalysed thermalization implies that $m_{S} \lae 10^{10-11} \GeV$.
Given that the inflaton mass scale in SUSY inflation models can naturally be  
$m_{S} \approx 10^{15-16} \GeV$, this can impose a significant constraint on inflation
 models 
compatible with $d=6$ AD baryogenesis. We also give the 
upper bound for the case $T = T_{R}$, corresponding to the case where 
the inflaton decay products instantaneously thermalize. This requires that 
$m_{S} \lae 10^{9-10} \GeV$ for $T_{R} \approx 1 \GeV$. These bounds are much
 weaker than would be expected from naive 
self-thermalization; comparing $m_{seed}$ with $m_{self}$ shows that 
self-thermalization would impose an upper bound smaller by a factor of more than
 $10^{3}$. For the case $T_{R} \approx 10^{3} \GeV$, 
the nucleosynthesis upper bound  requires that $m_{S} \lae 10^{12-13} \GeV$, 
which is less than would typically be expected in many inflation models, but 
which could nevertheless be satisfied with some moderately small couplings. For the case
 $T_{R} \approx 10^{3} \GeV$ we have also calculated
bounds for $T \approx 10^{2} \GeV$, corresponding to thermalization before the
 electroweak phase transition temperature, which is necessary for 
the existence of an electroweak phase transition. We see that 
this imposes quite a strong upper bound on the inflaton mass, $m_{S} \lae 
10^{11} \GeV$. Finally, instantaneous thermalization for $T_{R} \approx 10^{3} \GeV$ requires that $m_{S} \lae 10^{10-11} \GeV$.
For the case where $T_{R}$ is of the order of the conventional thermal gravitino 
upper bound, $T_{R} \approx 10^{8} \GeV$, the nucleosynthesis bound requires that
 $m_{S} \lae 10^{16-17} \GeV$, which is easily (although not necessarily trivially) 
satisfied in inflation models such as D-term inflation. The electroweak transition bound
 requires that 
$m_{S} \lae 10^{14-15} \GeV$. Again, although this can be satisfied in many inflation
 models, it is nevertheless of the same order of magnitude as the inflaton mass expected
 in D-term inflation models, for example. Therefore it is non-trivial to have
 thermalization before the electroweak phase transition temperature, even in realistic
 inflation models with relatively high reheating temperatures! Finally, instantaneous
 thermalization for $T_{R} \approx 10^{8} \GeV$ occurs only if  $m_{S} \lae 10^{12-13} \GeV$. 

     This shows that the assumption of a thermalized background following inflation,
 and in particular the existence of an electroweak phase transition, {\it strongly}
 depends upon the inflaton mass. In many inflation models it is likely that the
 relativistic particles of the radiation background will only thermalize at a relatively low
 temperature. Although for larger reheating temperatures it is quite easy to have
 thermalization before nucleosynthesis, it is not so clear that thermalization
 will occur before the epoch of  the electroweak phase transition. For lower reheating
 temperatures, thermalization 
before nucleosynthesis can impose significant bounds on the inflaton mass and so 
on the model of inflation. 

      Finally, we comment on some assumptions made in this discussion. We have
 assumed throughout that the inflation decay products are massless. Of course, only 
the photons and possibly the gluons and neutrinos can be treated as massless
 throughout. However, this is sufficient for our discussion of thermalization, since once
 these particles thermalize they will serve as a thermalized background which can
 thermalize all the other particles. In addition, we have also assumed that the only
 source of radiation is the single decaying inflaton field. In fact, shortly after the end of
 inflation, there may be other sources of radiation which could contribute to the
 thermalized seed leading to catalysed thermalization. This could provide a much larger
 number of seed particles, relaxing the upper bound from $m_{seed}$.
For example, in D-term inflation models, in addition to the inflaton there are the fields
 $\psi_{+,\;-}$ responsible for hybrid inflation \cite{dti}. The 
rapid decay of the $\psi_{-}$ field (which has a mass typically of the same magintude
 as the inflaton mass \cite{dti,em}) releases roughly the same energy in a time $\delta t
 \approx H_{I}^{-1}$ as that stored in the inflaton field at the end of inflation.
 Therefore the number of lowest energy particles in the spectrum, corresponding to
 those produced at $H \approx H_{I}$, is enhanced by a factor $H_{I}/H_{R}$. The
 result is that the $m_{seed}$ upper bound is increased by a factor 
$(H_{I}/H_{R})^{1/3}$, which is greater than $10^{5}$ for $H_{I} \approx
10^{13} \GeV$ and $T_{R} \lae 10^{8} \GeV$. Thus the bounds on the inflaton
mass in hybrid inflation models are likely to be much weaker than in models with a
single inflaton field. 
\section{Conclusions} 

             We have considered the constraints following from the requirement that the 
relativistic decay products from inflaton decay thermalize. We have shown that the 
low energy decay products from inflaton decays occuring shortly after the end of
 inflation play a vital role in the thermalization of the whole energy spectrum, a process 
we refer to as catalysed thermalization. For the case where a single decaying inflaton field is
 the source of the thermal energy,  requiring that the decay products  
do not thermalize before the temperature of the thermal gravitino upper bound 
allows the upper bound on the "reheating temperature" in SUSY inflation models to be
 increased from 
$10^{8} \GeV$ to $10^{10-12} \GeV$ in realistic inflation models. Requiring 
that thermalization occurs before to onset of nucleosynthesis can impose tight upper
 bounds on the inflaton mass for low reheating temperatures, such as may be required
 by Affleck-Dine baryogenesis or non-thermal gravitino production. In addition, even
 in realistic inflation models with a relatively high reheating temperature, it is quite
 possible that the relativistic background will not thermalize before the temperature of
 the electroweak phase transition. In hybrid inflation models, such as D-term inflation,
 the thermal energy due to the decay of the hybrid inflation fields at the end of inflation
 can provide a seed of thermalized particles which makes subsequent thermalization of
 the inflaton's decay products much more efficient, relaxing the upper bounds on the
 inflaton mass. The lesson from all this is that in many inflation models the thermal history of
 the Universe is likely to be quite different from that naively assumed on the basis of 
instantaneous thermalization of the inflaton's decay products. 

\subsection*{Acknowledgements}   Support from the PPARC, UK, is acknowledged.

\end{document}